\documentclass[manuscript]{acmart}

\AtBeginDocument{%
  \providecommand\BibTeX{{%
    \normalfont B\kern-0.5em{\scshape i\kern-0.25em b}\kern-0.8em\TeX}}}

\setcopyright{acmcopyright}
\copyrightyear{2018}
\acmYear{2018}
\acmDOI{XXXXXXX.XXXXXXX}

\acmConference[]{}{}
\begin{document}
%%
%% The "title" command has an optional parameter,
%% allowing the author to define a "short title" to be used in page headers.
\title{Co-Design with Myself: A Brain-Computer Interface Design Tool that Predicts Live Emotion to Enhance Metacognitive Monitoring of Designers}

%%
%% The "author" command and its associated commands are used to define
%% the authors and their affiliations.
%% Of note is the shared affiliation of the first two authors, and the
%% "authornote" and "authornotemark" commands
%% used to denote shared contribution to the research.
\author{Qi Yang}
\email{qy244@cornell.edu}
\orcid{0000-0002-8063-266X}
\affiliation{%
  \institution{Human Centered Design, Cornell University}
  \streetaddress{2427 Martha Van Rensselaer Hall}
  \city{Ithaca}
  \state{New York}
  \country{USA}
  \postcode{14853}
}

\author{Shuo Feng}
\affiliation{%
  \institution{College of Architecture, Art, and Planning, Cornell University}
  \streetaddress{129 Sibley Dome}
  \city{Ithaca}
  \country{USA}}
\email{sf522@cornell.edu}

\author{Tianlin Zhao}
\affiliation{%
  \institution{Department of Computer Science, Cornell University}
  \city{Ithaca}
  \country{USA}
\email{tz68@cornell.edu}
}

\author{Saleh Kalantari}
\affiliation{%
  \institution{Human Centered Design, Cornell University}
  \city{New York}
  \country{USA}}
\email{sk3268@cornell.edu}

%%
%% By default, the full list of authors will be used in the page
%% headers. Often, this list is too long, and will overlap
%% other information printed in the page headers. This command allows
%% the author to define a more concise list
%% of authors' names for this purpose.
\renewcommand{\shortauthors}{Qi, et al.}

%%
%% The abstract is a short summary of the work to be presented in the
%% article.
\begin{abstract}
Intuition, metacognition, and subjective uncertainty interact in complex ways to shape the creative design process. Design intuition, a designer's innate ability to generate creative ideas and solutions based on implicit knowledge and experience, is often evaluated and refined through metacognitive monitoring. This self-awareness and management of cognitive processes can be triggered by subjective uncertainty, reflecting the designer's self-assessed confidence in their decisions. Despite their significance, few creativity support tools have targeted the enhancement of these intertwined components using biofeedback, particularly the affect associated with these processes. In this study, we introduce "Multi-Self," a BCI-VR design tool designed to amplify metacognitive monitoring in architectural design. Multi-Self evaluates designers' affect (valence and arousal) to their work, providing real-time, visual biofeedback. A proof-of-concept pilot study with 24 participants assessed its feasibility. While feedback accuracy responses were mixed, most participants found the tool useful, reporting that it sparked metacognitive monitoring, encouraged exploration of the design space, and helped modulate subjective uncertainty. 
\end{abstract}

%%
%% The code below is generated by the tool at http://dl.acm.org/ccs.cfm.
%% Please copy and paste the code instead of the example below.
%%


%%
%% Keywords. The author(s) should pick words that accurately describe
%% the work being presented. Separate the keywords with commas.

%% A "teaser" image appears between the author and affiliation
%% information and the body of the document, and typically spans the
%% page.
\begin{teaserfigure}
  \includegraphics[width=\textwidth]{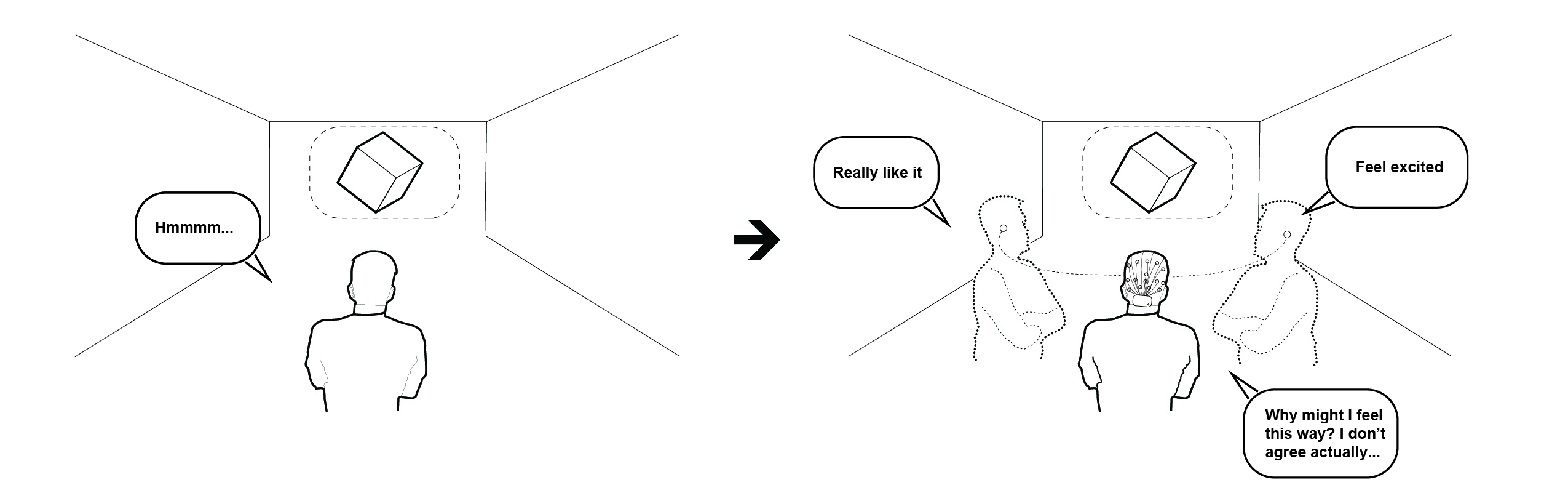}
  \caption{Concept Diagram of Co-design with Myself Using The “Multi-Self” Design Tool.}
  \Description{concept image}
  \label{fig:teaser}
\end{teaserfigure}

%%
%% This command processes the author and affiliation and title
%% information and builds the first part of the formatted document.
\maketitle

\section{Introduction}
Design is regarded as a non-linear and iterative process of divergent and convergent thinking. During the divergent thinking processes, designers frame the design space and explore as many design options as possible, followed by design evaluation and synthesis in the convergent thinking process \cite{cross2023design}. The exploration stage is characterized by high uncertainty, complicating the anticipation of the effectiveness of initial decisions given the potential complexity they might introduce, particularly in wicked design problems. Designers sometimes rely on intuition to make judgments, but the lack of self-reflection and awareness in their responses to different design options make many designers second guess themselves \cite{schut2020towards,mclaren2008exploring, faste2017intuition}. When designers ask themselves “Am I on the right track” or “If I have found the right design”, the answers could emerge from their intuitive confidence rather than rational evaluations. Sometimes, designers already have the answer but they hesitate to move forward. If lacking confidence, excessive hesitation could discourage designers from thoroughly exploring the design space and limit their creativity. Strong intuition and being able to maintain awareness of their self-confidence in early design stages are important for creative design practices \cite{nickerson1999enhancing}.

Intuition, self-awareness, and feelings of confidence are entangled constructs in the design context. Despite the complex nature of intuition, it is described as a combination of unconscious evaluative processes and conscious reflective acts in the design practice \cite{faste2017intuition}. The conscious reflective practice, also known as self-awareness or metacognitive monitoring, describes the process of “thinking about thinking” and consciously adopting effective mental strategies \cite{ackerman2017meta}. Indeed, being able to identify the strength and weaknesses of one’s own ideas and be aware of the rationales behind their decision-making will improve better idea evaluation and selection \cite{puente2021metacognitive}. Feelings of confidence about design decisions, named epistemic uncertainty by Ball and Christensen, is also closely related to design evaluation \cite{christensen2018fluctuating}. According to this outlook, fluctuations in epistemic uncertainty may be an important trigger that prompts designers to alter or reinforce their mental approach to a design challenge. Too much uncertainty may limit the creative process by creating insecurity, apprehension, and creative paralysis \cite{ball2019advancing}. Too little uncertainty during the early stages of design may contribute to design fixation. Some epistemic uncertainty can lead to deliberate counterarguments, initiate creative reconsideration, and potentially push a design forward \cite{paletz2017dynamics,christensen2018fluctuating}. In theory, metacognitive monitoring plays a crucial role in modulating and keeping epistemic uncertainty within effective parameters, thereby avoiding the extremes of design fixation or creative paralysis \cite{carlson2020design, crilly2017next, mclaren2008exploring}. As demonstrated, self-awareness and the ability to perceive accurate uncertainty and modulate a feeling of confidence in design are critical parts of strong design intuition. Building up strong intuition requires years of strategic practice. Even so, maintaining awareness of confidence is still challenging because internal mental states are difficult to perceive.  

Recognizing the importance of intuitive feelings towards design, our motivation is to explore the possibility of a design tool that could modulate designers’ subjective uncertainty to enhance metacognitive monitoring through affective computing and biofeedback, specifically in an architectural design context. Architectural design, given its complexity and multifaceted nature, is challenging to evaluate during early design stages. This design discipline involves careful consideration of aesthetics, functionality, sustainability, and sociocultural factors. Such characteristics of architectural design make it a particularly suitable platform for testing our tool. \cite{oxman2006theory}. Affect is closely related to intuition, metacognition, and uncertainty. One common model of affect theory describes emotions that emerge in response to environmental stimuli as a blend of three independent factors: (1) pleasure, also known as valence, which is an overall feelings of “positivity” or “negativity” towards something; (2) arousal, which describes the degree of “excitement”; and (3) dominance, which refers to feeling “unrestricted” or “restricted” in an environment \cite{mehrabian1974basic}. Affects such as “gut feelings” are considered to be foundations of intuition \cite{epstein2010demystifying}. Previous research has shown that metacognitive experiences are generally associated with improved affect, especially with feelings of pleasure \cite{efklides2006metacognition, efklides2011interactions}. Uncertainty is usually associated with negative affect, but it can be positive when present in limited amounts in creative contexts \cite{anderson2019relationship}. Knowing the close relationship between affect and intuition, feelings of confidence, and metacognition, we envision a design tool that predicts if the user feels excited or positive towards the current design. More importantly, the tool will visualize those predictions to users in real time to help them perceive their internal states in a third-person perspective, in hopes of increasing self-awareness, modulating feelings of confidence, and improving intuition (Figure \ref{fig:teaser}). 

Prior work has used a variety of means to identify affect. One approach is grounded in evaluating facial expressions and gestures, which can be done computationally via emotion recognition algorithms. Such evaluations, however, are based on secondary expressions of affect, which can vary quite a bit in different cultural, social, and individual contexts \cite{elfenbein2002universality}. Most importantly for our purposes, design is often a solitary process and when working alone many designers may not register their feelings via facial expressions. Another method that has been used to identify some forms of affect is electrodermal activity (EDA), but this physiological reaction is lacking in nuance and does not change fast enough to capture the rapidly changing affect responses that may occur during the design process \cite{caruelle2019use}. Using EEG measurements provides a much greater level of detail and directly evaluates brain responses. Previous studies have shown that affective states can be measured via EEG, including affective responses to architectural environments  \cite{banaei2020emotional, coburn2020psychological, shemesh2021neurocognitive, vartanian2015architectural}. Additional studies have classified pleasure, arousal, and dominance in numerous contexts and activities, providing benchmarks of accuracy that can be used to evaluate our BCI performance  \cite{galvao2021predicting, hu2017eeg, khosrowabadi2010classification, wichakam2014evaluation, marin2018affective, suhaimi2018modeling}.

In sum, the current project provides several important contributions to human-computer interaction research and to the development of creativity-support tools:  
\begin{itemize}
\item We present a closed-loop BCI design tool that identifies and visualizes users’ real-time affect levels based on their EEG data. 
\item We have examined the BCI tool’s feasibility and usability by conducting user testing.
\item The research qualitatively evaluates the potential for real-time affect awareness to modulate feelings of confidence, stimulate metacognitive reactions, and support creativity for designers.
\item We discuss the potential applications of this work and future avenues for affective computing and biofeedback in supporting creative design processes.
\end{itemize}

\begin{figure}[h]
  \includegraphics[width=\textwidth]{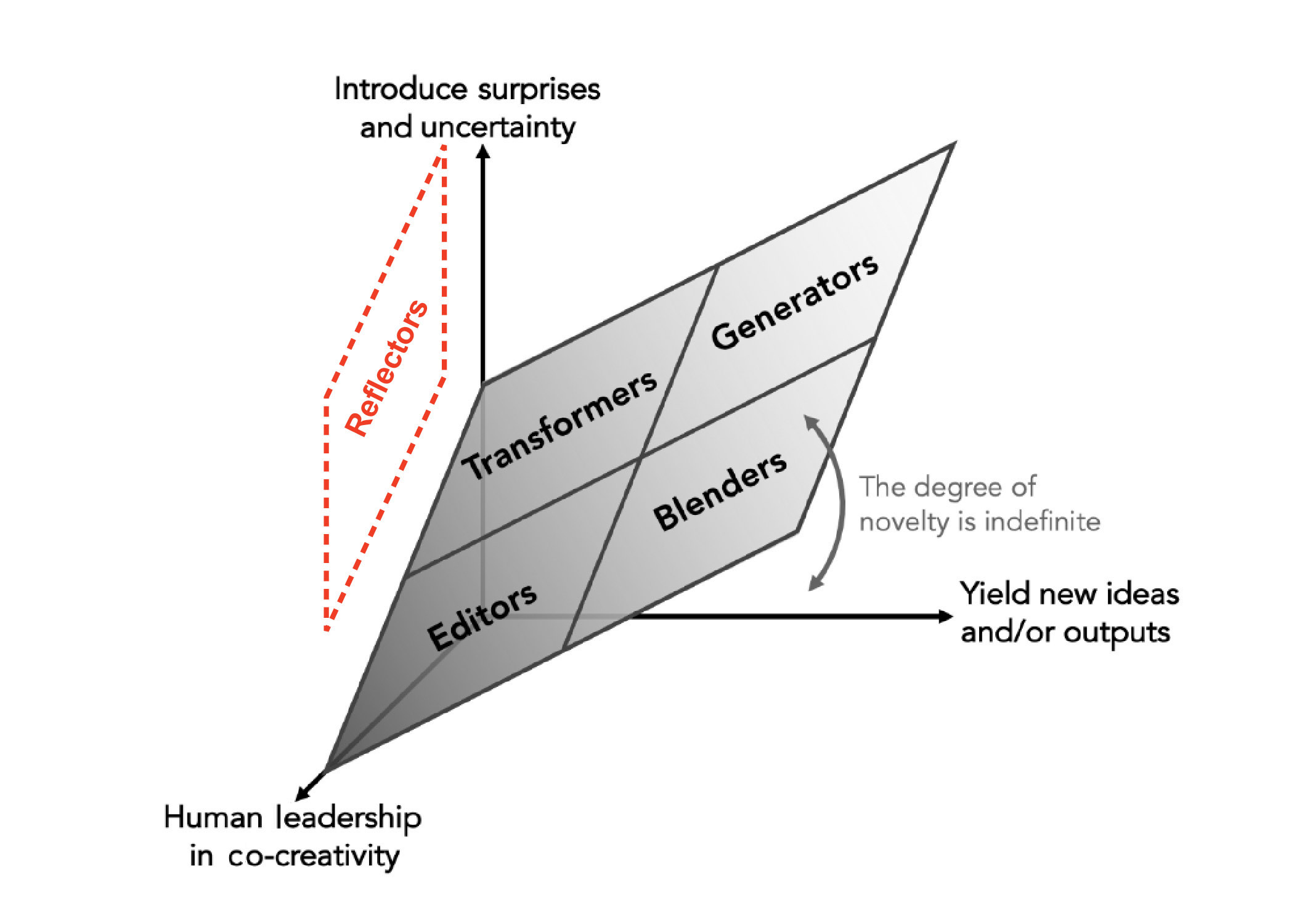}
  \caption{A New “Reflectors” Type of AI Co-creative Tool.}
  \Description{reflector image}
  \label{fig:reflector}
\end{figure}

\section{Related Works}

Here we reviewed related works on EEG-based emotion detection, devices or tools that provide biofeedback to users, and human-AI co-creative tools.

\subsection{Emotion Detection Based on BCI-VR}
Given its wide range of applications, automatic emotion classification is gaining the attention of scholars in HCI fields. Recent improvement in consumer-grade wearable EEG devices captures interest among the HCI community in using EEG for emotion classification \cite{suhaimi2020eeg}. Particularly focused on arousal and valence affective response, previous studies have used the EEG classification approach in the context of using music stimuli \cite{nawaz2018recognition}, music videos \cite{shahnaz2016emotion, byun2017feature, xu2018eeg, wu2017estimation, ullah2019internal, dabas2018emotion}, and video clips \cite{song2018eeg, liu2017real, kimmatkar2018human, nie2011eeg, terasawa2017tracking, li2019classification}. The immersive virtual reality (VR)’s ability to evoke emotion gives VR more interest to be used as a tool in emotion detection in general \cite{zhang2017affective, kim2018effect, hidaka2017preliminary, fan2017eeg, horvat2018assessing, guo2019effect}. In fact, the self-reported intensity of emotion was found to be significantly greater in immersive VR compared to similar content in non-immersive virtual environments \cite{chirico2017effectiveness}. 
Architectural design is unique compared to other design fields due to its focus on human responses to spatial environments \cite{djebbara2019sensorimotor}. In a relevant VR study, Marín-Morales et al. \cite{marin2018affective} used an EEG classification approach and reached an accuracy of 75.00\% along the arousal dimension and 71.21\% along the valence dimension (chance level: 58\%) in the classification of four alternative virtual rooms. The same study was one of the first studies that developed an emotion recognition system using a set of immersive VR as a stimulus elicitation and discussed the application of the approach in architecture. Other studies employed the BCI-VR system to evaluate participants' EEG responses to different architectural designs \cite{kalantari2021comparing, darfler2022eeg, kalantari2022evaluating, cruz2022eeg, rounds2020using, koh2020electroencephalography}. Therefore, we posit that exploring a more efficient integration of Brain-Computer Interface (BCI) and Virtual Reality (VR) in architectural design could facilitate seamless spatial feedback.

\subsection{Biofeedback to Myself}

Several studies have been done to help users be aware of their inner states through different modalities of feedback. One study developed Self-Interface, an on-body device attaching to the spine that helped users perceive physiological signals by translating them to haptic feedback \cite{haghighi2020self}. Similarly, another study helped users to perceive their pupil dilation that are associated with different cognitive tasks by making pupil dilation audible. Their findings suggested most users were able to associate cognitive activities with the sounds, which elicited metacognitive awareness \cite{de2018augmented}. Besides haptic and audial feedback, a third study translated heart rate into shape-changing displays, which brought awareness of users’ physiological states through visual feedback \cite{yu2016livingsurface}. The Inner Garden project created a garden where mental states identified by EEG and breath sensors were translated into weather and day duration of the world. The device was meant to encourage self-reflection and foster mindfulness \cite{roo2016inner}. With a similar aesthetic pursuit, this study explored the system structure of visualizing fMRI data aesthetically in an immersive environment \cite{thompson2009allobrain}. The majority of the studies focused more on developing the feedback system and less on user experiences in the context of an application. Nonetheless, those studies offer creative ideas about how biofeedback could be perceivable and suggest how perceiving biofeedback could benefit metacognitive awareness. Other applications of biofeedback have been explored such as modulating breathing \cite{prpa2020inhaling}, enhancing storytelling, promoting communication \cite{semertzidis2020neo, frey2020physiologically} and gaming \cite{huang2015heartbeat, nacke2011biofeedback}. One interesting study in the sleep facilitation context discussed design strategies that we found adaptable to our study in the design context \cite{semertzidis2019towards}. The study pointed out the importance of motivating exploration, promoting responsiveness of BCI, and facilitating self-expression. Our study shared the same goal to increase user awareness in a different context. Though we found few projects have specifically explored biofeedback’s potential in supporting the creative design process.

\subsection{Co-create and Co-Evaluate with an Agent}

Creativity, fundamentally a complex cognitive process, is defined as the ability to generate ideas or products that are both novel and valuable \cite{sternberg2003wisdom}. It's a skill that transcends the traditional boundaries of art or design, playing a crucial role across various fields, from science to business. To foster such creativity, a wide range of tools, known as Creativity Support Tools (CSTs), have been developed. CSTs are technologies designed to enhance human creative thinking and activities by offering innovative ways to seek, manipulate, and represent information and ideas \cite{shneiderman2006creativity}.

\par
Davis and colleagues \cite{davis2015enactive} proposed an intriguing framework comprised of three models outlining how computational systems could engage in co-creation alongside humans. The first model aims to boost creativity by enabling users to more types of execution. The second model operates by autonomously generating creative design options. The third model, named "computer colleagues", focuses on "collaborating with humans in continuous real-time improvisation to enrich the creative process." The objective of 'computer colleagues' is to foster creative products through iterative interactive negotiations between multiple parties.

Interestingly, most creativity support tools align with the first two categories. For example, a tool that facilitates users to juxtapose images with texts, enabling exploration of broader ideas and relationships \cite{kerne2008combinformation}. Another system automatically generates diverse alternative design options, assisting users in designing icons \cite{zhao2020iconate}. Within the architectural design, a study has developed a digital-twin tool. This tool allows users physically prototype their designs and simultaneously analyze the outcomes digitally. \cite{kalantari2022developing}. Using BCI, some other studies enable users to generate and manipulate geometries by thinking \cite{yang2023mindopen, yang2021mindsculpt, shankar2014human}. Regarding the third category, a prevalent approach involves developing intelligent creative agents with human-like cognition derived from rules or data, anticipating creative outcomes to emerge from agreements or conflicts between the two parties \cite{karimi2019creative, louie2020novice}.

Beyond these three categories, some projects, such as DesignEye, have aimed to support the design process by augmenting visual perception and providing analytical maps (for instance, saliency and visual clutter maps), which are beyond the naked eye's capability to discern \cite{rosenholtz2011predictions}. In a similar vein, Machinoia aimed to visualize users' potential attitudes at different times simultaneously \cite{pataranutaporn2021machinoia}.

Inspired by these advances, our project aspires to create a 'computer colleague' while also sharing the goals of augmenting users' perception and cognition with the aforementioned projects. Diverging from previous endeavors, our investigation centers around a unique proposition: what if the 'computer colleague' was partially the user themselves? We aim to explore whether this "co-design with myself" type of interaction triggers enhanced metacognitive monitoring processes by visualizing typically unseen biofeedback.

In another recent review article, Hsing-Chi Hwang \cite{hwang2022too} argued that AI-based co-creative tools can be divided into four types: “Editors” that enable specific design execution, “Transformers” that change design content to different forms, “Blenders” that mix creative elements together, and “Generators” that produce entirely novel creative products. We find it notable that this categorization schema does not include tools enhancing users' cognitive abilities such as metacognitive monitoring that we envisioned for the current project. Thus, we propose a fifth type of co-creative tool, “Reflectors.” They don’t generate, but instead, focus on empowering designers’ self-awareness and enhancing their intuition during the process of confronting a design problem (Figure \ref{fig:reflector}). Indeed, human-AI co-creative process will enrich the design space and dynamics of design iteration. We think it is equally important to develop a co-evaluating agent that helps users select their creations in a creative way.

\section{Method}

We present the BCI design tool “Multi-Self” which aids architectural design processes. It was developed with two primary parts. The first was an application that allowed designers to create an interior lobby space in virtual reality. We chose the task of designing a lobby because it was a very common design problem. A lobby’s function and range of potential forms were quite flexible compared to other architectural features \cite{ching2018interior}. 

To minimize motor actions and acquire clean EEG data of affect during the design selection, we developed a mixed-modality control function for this lobby design process, which combined head pointing and key presses. Users were able to focus on a design option through eye movements and then press buttons to adjust and confirm the selection. In addition to providing greater ease/fluidity in the design process, this approach helped to reduce large motion artifacts that can limit the effectiveness of the EEG classification.

\begin{figure}[h!]
  \includegraphics[width=\textwidth]{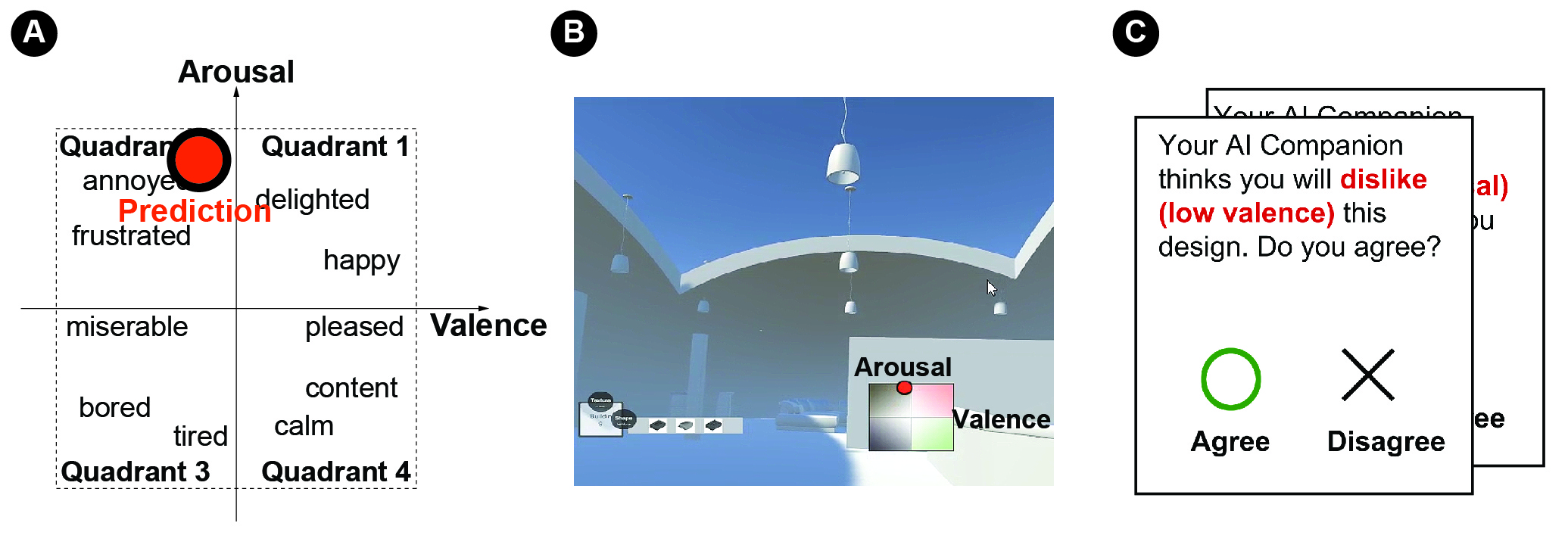}
  \caption{(A) Using the Emotion Coordinate System to Display the Results of EEG Classification; (B) Real-Time Visual Feedback in VR from Multi-Self; (C) Neurological Feedback and Pop-up Panels from "Multi-Self" in VR during the Validation Session.}
  \Description{Classify}
  \label{fig:Classify}
\end{figure}

The second primary component of Multi-Self was a BCI algorithm that categorized users’ affective responses to environmental stimulus (refer to section \ref{sec:stimuli}) based on their EEG data, and provided real-time biofeedback about these responses within the VR display. The form of this feedback was based on the emotion coordinate system developed by Lang \cite{lang1995emotion}, which is widely used for affective computing \cite{koelstra2011deap, soleymani2015analysis, lin2010eeg}. The first quadrant displayed high-arousal and high-valence states (emotions generally described as “joyful” or “excited”). The second quadrant displayed high-arousal and low-valence states (“fearful” or “enraged”). The third quadrant displayed low-arousal and low-valence states (“boring” or “sad”), and the fourth quadrant displayed low-arousal and high-valence states (“relaxed” or “calm”). During the design process, users’ live emotional feedback was visualized in the corner of the VR display as a moving colored point on this coordinate graph (Figure \ref{fig:Classify}).

\begin{figure}[h!]
  \includegraphics[width=\textwidth]{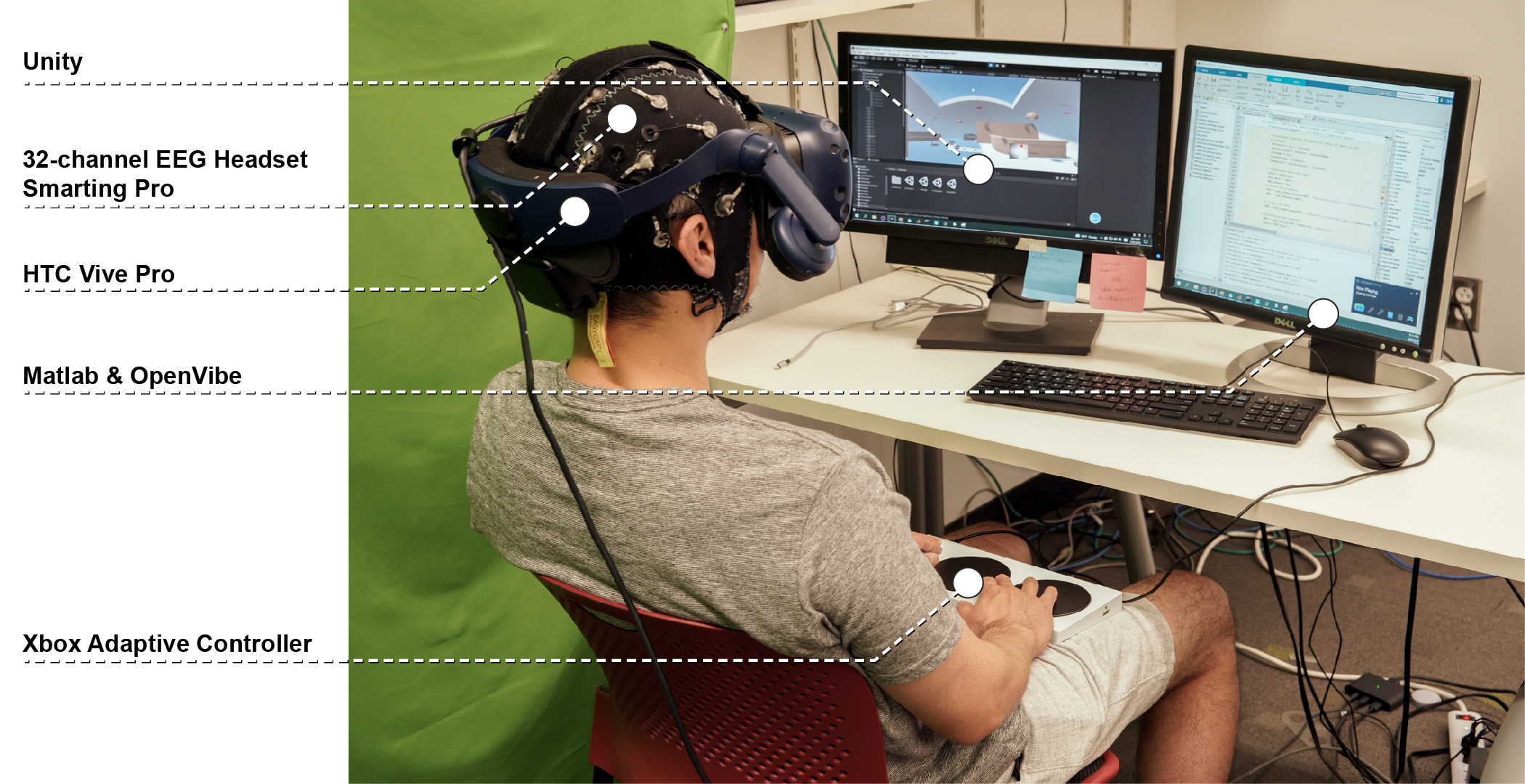}
  \caption{Experiment Setting of the Augmented Co-Design Process.}
  \Description{Experiment}
  \label{fig:experimentsetting}
\end{figure}

\subsection{Environmental Stimuli Used for Training the BCI}
\label{sec:stimuli}
\begin{figure}[h]
  \includegraphics[width=\textwidth]{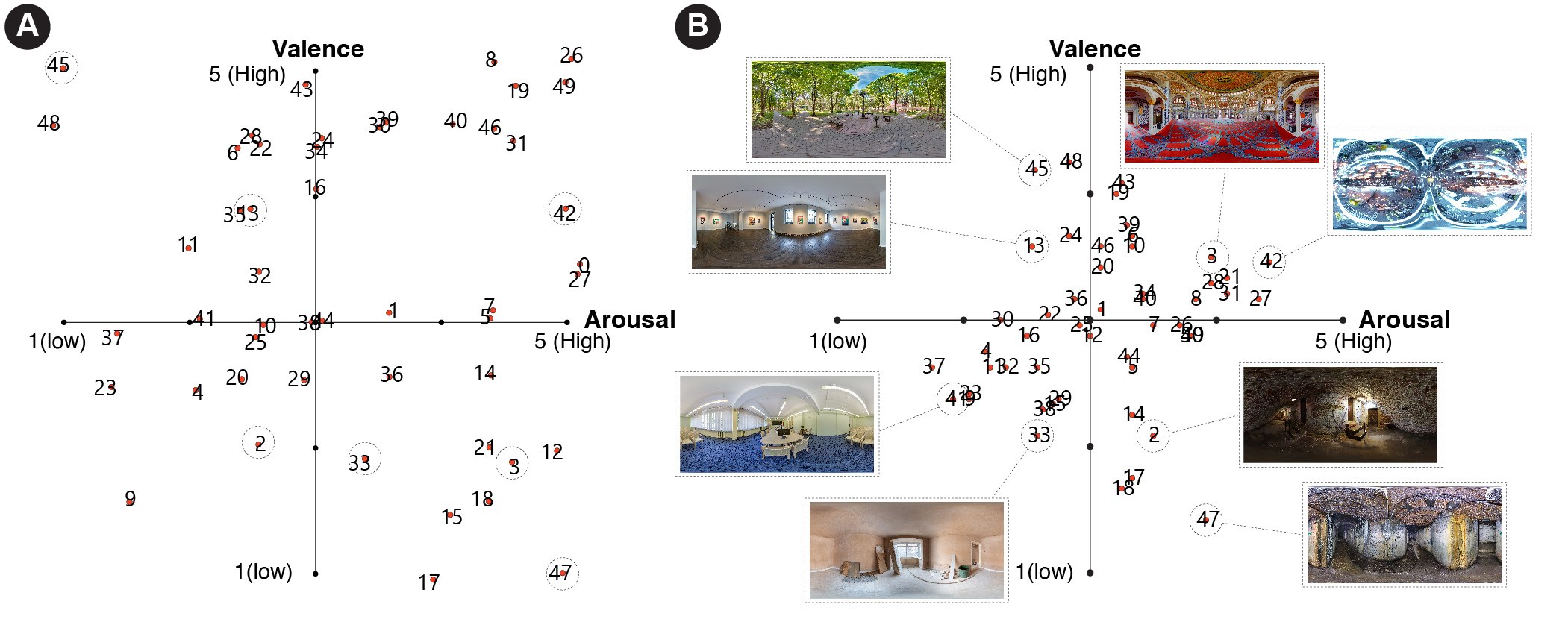}
  \caption{The Emotion Ratings of Stimuli by the Judges and the Participants: (A) Average Emotion Ratings of Two Judges; \emph{Dot locations were slightly dodged to make overlapping dots visible.} (B) Average Emotion Ratings of 24 Participants. \emph{Dotted circles highlight some stimuli for comparison.}}
  \Description{Protocol}
  \label{fig:stimuli}
\end{figure}

The most common environmental datasets used for emotional response analysis in prior research, such as the DEAP dataset \cite{koelstra2011deap}, were in the format of two-dimensional videos. We were able to find only two existing datasets offering immersive 3D videos for this purpose. One of them included outdoor human and animal activities and was thus unsuited for architectural design analysis \cite{miller2020personal}. Another small existing 3D dataset did include indoor environmental stimuli, but this amounted to only four samples, which was insufficient for our machine learning needs \cite{marin2018affective}. One recent study systematically developed immersive stimuli to elicit five emotions. However, they were not available when the study was conducted \cite{dozio2022design}. Therefore, we curated our own immersive 3D environmental stimuli to stimulate affective responses from the participants. Given that we requested participants to annotate their emotional reactions to various architectural environments, we tried to provide a selection of environments that could elicit a diverse range of emotions. This approach aimed to prevent a situation where certain emotional responses were overly represented in the labels. In the first step of this process, we download one hundred 360-degree panoramic images from the internet. Some of them were images of real-world environments and others were renderings of imagined spaces. To remove stimuli that would be unlikely to provoke an affective response, two judges with over 5 years of architectural design experience were asked to view all of the downloaded images via the same HTC-Vive headset used in the experiment and to report their arousal and valence levels related to each image using the SAM scale. We removed images that received “moderate” scores from both judges on each axis (scores of 2 or 4, out of a 1–5 range). The final selection of 50 images was made with the goal of obtaining a balance between “high” (5), “neutral” (3), and “low” (1) scores on each axis, to ensure that each quadrant on the Emotion Coordinate System was reasonably represented. Due to one operation error from the experimenter, the first stimulus was systematically not presented to all the participants. Eventually, participants experienced 49 stimuli. The distribution of judges' ratings and the average ratings of 24 participants were shown in Figure \ref{fig:stimuli}. There is a correlation of 0.713 between the participants' average ratings and the researchers' ratings for arousal and a similarly close correlation of 0.709 for valence.

\subsection{Design the Design Space}
\label{sec:design space}
We broke down the lobby design process into four components, aiming to facilitate users from diverse design backgrounds to participate in the design process: the \textit{Envelope} that determined the overall geometry and windows of the room, the \textit{Layout} of furniture arrangements, the \textit{Fixtures} that in this case were limited to lighting, and the \textit{Color} palette for the various room components. 

\begin{figure}
  \includegraphics[width=\textwidth]{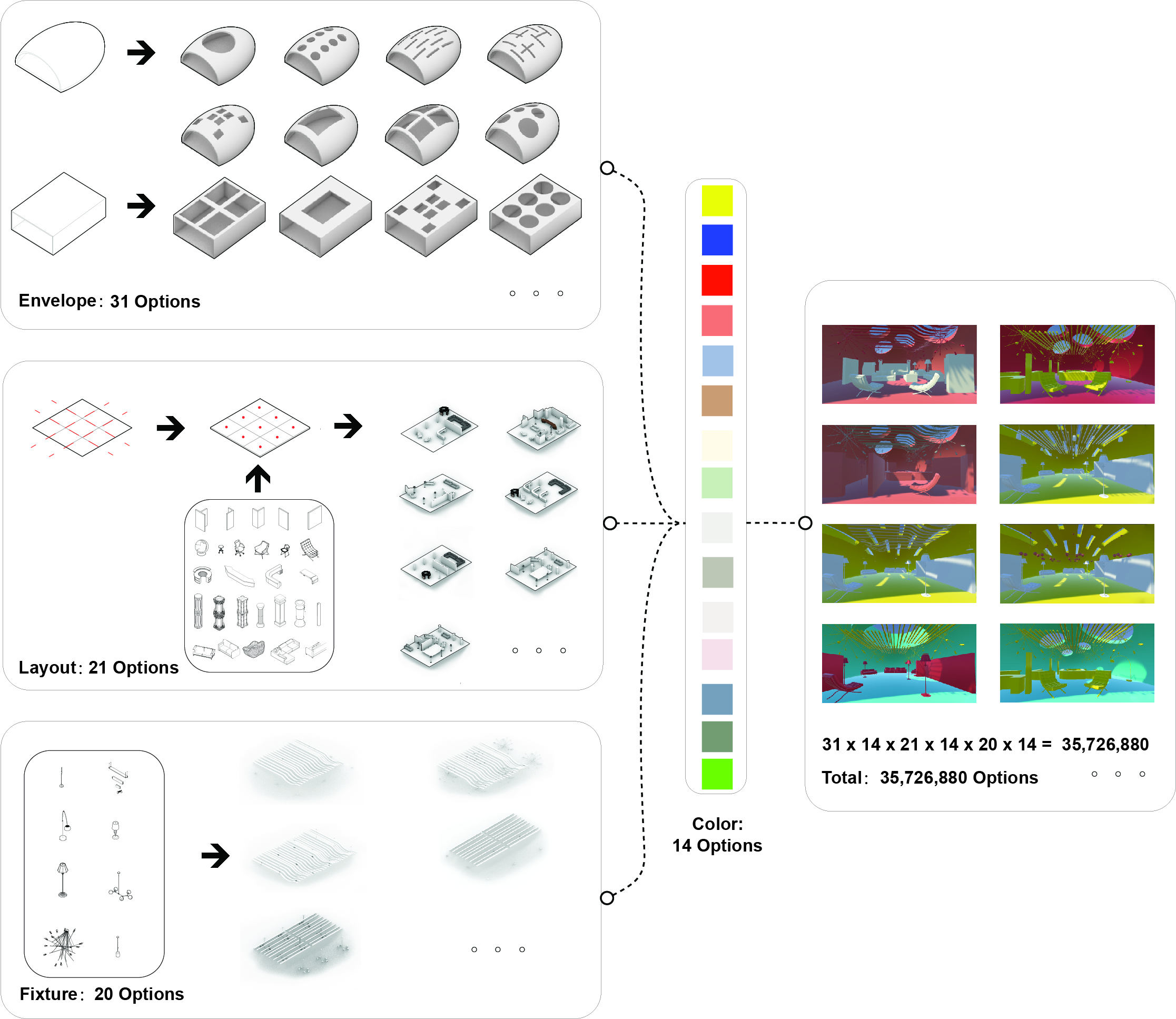}
  \caption{Designing a Lobby Space by Combining Envelope, Layout, Fixtures, and Color.}
  \Description{Generation}
  \label{fig:Generation}
\end{figure}

We created 31 envelope options, categorized into rectangular and arched forms, accommodating a range of skylight sizes. For the layout, we first partitioned the lobby space into nine distinct sections, systematically positioning selected building elements within each region. This process generated 500 unique layout options using building elements sourced from the Thingiverse website \cite{Thingiverse}. After careful examination, we narrowed down to 21 distinct layouts with diverse designer input. Regarding the fixture aspect, we selected lighting fixture models, also from Thingiverse, for their broad compatibility, ensuring a balanced mix of floor lamps and pendant lights. We also modeled twenty varied ceiling fixture options with different forms and densities. We combined the ceilings with the lighting fixtures considering factors such as quantity, position, and type of light sources. Finally, our selection process included a palette of 14 colors, each differing in hue, saturation, and brightness. The free combination in each of the categories yields 35,726,880 potential solutions in the design space (Figure \ref{fig:Generation}).

\subsection{Hardware and Software Setup}
\begin{figure}[h]
  \includegraphics[width=\textwidth]{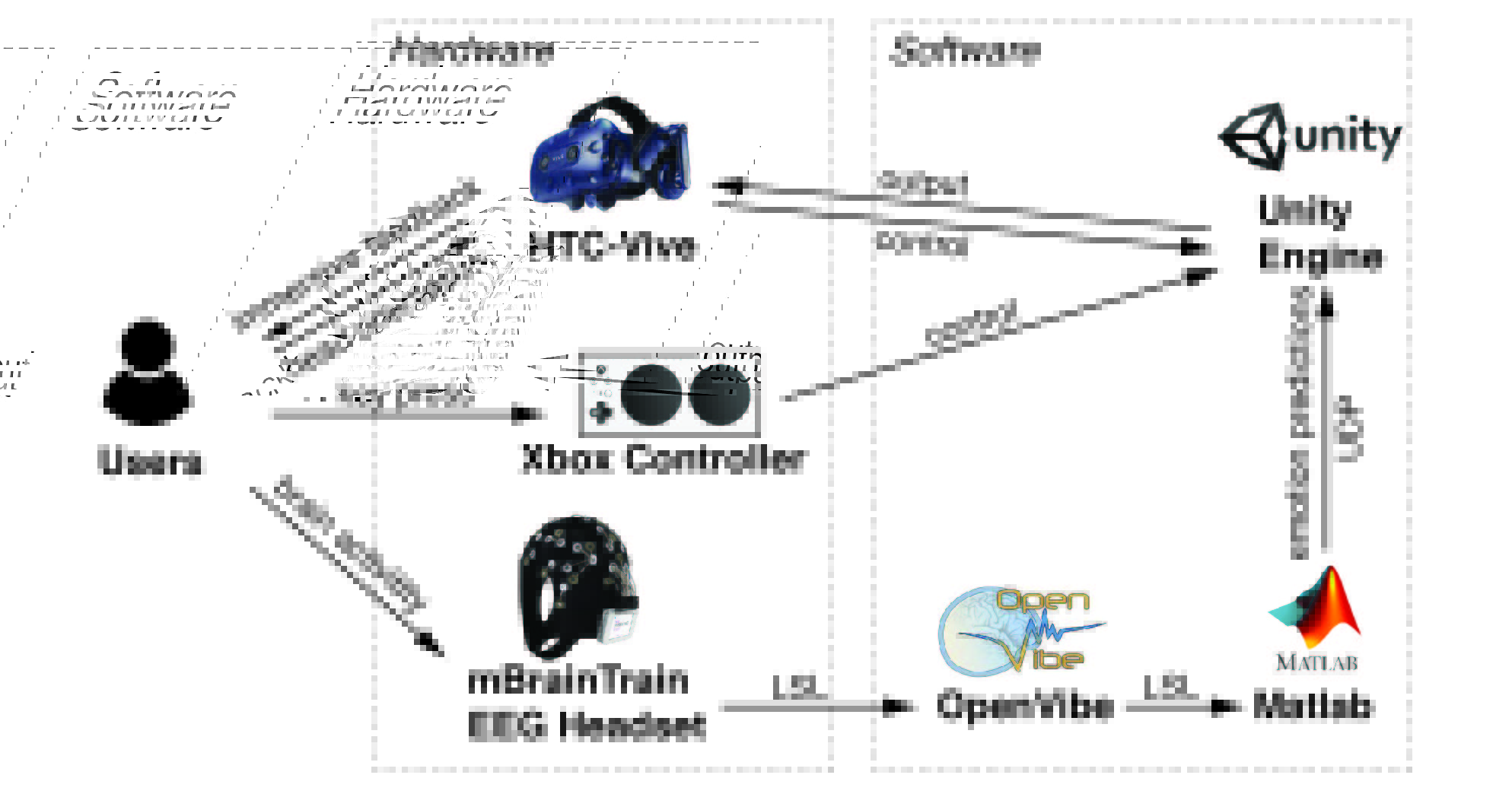}
  \caption{System Diagram of the Multi-Self Design Tool.}
  \Description{System Diagram}
  \label{fig:system}
\end{figure}

The BCI design tool incorporated an EEG headset, a VR headset, an Xbox adaptive controller, and a combination of commercial and open-source software (Figure \ref{fig:system}). EEG data were acquired from a non-invasive, 32-channel, gel-based mBrainTrain Smarting Pro (mBrainTrain LLC., Belgrade, Serbia). Channels located at Fp1, Fp2, F7, F3, Fz, F4, F8, FT9, FC5, FC1, FC2, FC6, FT10, T7, C3, Cz, C4, T8, TP9, CP5, CP1, CP2, CP6, TP10, P7, P3, Pz, P4, P8, POz, O1, O2 based on the International 10-20 system. An HTC Vive head-mounted display was worn on top of the EEG headset (Figure \ref{fig:experimentsetting}). EEG data collected during the “offline” training session (offline: data analysis happens after data collection; online: data analysis in real-time) for each participant were preprocessed using the open-source EEGLab toolbox in commercial MATLAB R2020b \cite{delorme2011eeglab}. MATLAB was also used to train the models for interpreting and categorizing participants’ neurological data. During the online testing sessions, the real-time EEG data was acquired by the open-source tool OpenVibe \cite{renard2010openvibe} and sent to MATLAB through the Lab Streaming Layer \cite{delorme2011eeglab}. The EEG classification results were then delivered from MATLAB to the commercial Unity3D virtual-reality display engine via the User Datagram Protocol (UDP).

\subsection{EEG Pre-processing Pipeline and Machine Learning}

EEG data from the training session and the online session followed the same pre-processing pipeline. The data went through an initial Artifact Subspace Reconstruction (ASR) procedure via the commercial mBrainTrain Streamer software (mBrainTrain LLC., Belgrade, Serbia). After this procedure, the EEG data stream was bandpassed to isolate the relevant frequency band (1–40Hz). Then, bad channels that were flat for more than 5 seconds or exceed a high-frequency noise standard deviation of 4 were removed and interpolated from surrounding channels. The EEG data were re-referenced based on the average band power from all channels. For each of the 32 channels, theta (4–7 Hz), alpha (7–15 Hz), and beta (15–30 Hz) band powers were extracted in 2-second time windows (with 0.5-second overlaps). We then used the Minimum Redundancy Maximum Relevancy (mRMR) algorithm \cite{peng2005feature} to select the alpha, beta, and theta band power features of the EEG data that best predicted differences in emotional responses. This algorithm involved an iterative loop in which 4 data segments are used as a training set and 1 was used for validating the prediction, eventually leading to the identification of the data features that gave the best predictive accuracy. The validation set was chronologically partitioned to avoid temporal correlation that would mistakenly improve the accuracy \cite{li2020perils}. 

We trained two 3-class, user-specific, linear Support Vector Machine (SVM) models to predict high, neutral, and low arousal, and high, neutral, or low valence. The use of a traditional machine learning classifier like the linear Support Vector Machine (SVM) is well-suited for the automated identification of pre-set EEG frequency band-power features. There is substantial precedent for using SVMs for EEG classification \cite{lotte2018review, wei2021comparison, lee2019comparative}, with studies showing that SVMs sometimes outperform other machine learning methods. Preliminary tests using training and validation sets from initial participants indicated that both a kernel-SVM with a second-degree polynomial kernel and a linear SVM had good performance. However, we chose to proceed with the more straightforward algorithm, the linear SVM, to enhance the potential for generalizability. The classification analysis was conducted using the fitcecoc function in MATLAB.

\subsection{Participants}
The pilot test was intended to provide preliminary feasibility and user-feedback data, rather than robust confirmation of scientific hypotheses. 24 participants with normal or corrected-to-normal vision were recruited through convenience sampling (word-of-mouth and announcement on departmental e-mail lists). Ten participants had an architectural design background. Their ages ranged from 19 to 36 (M=22.9, SD=4.86). 14 participants reported as female and 10 participants reported as male. Three participants indicated that they had previously used BCIs on rare occasions, and 21 participants indicated that they had never used BCIs. Four participants reported having more than 8 years of design experience; Six participants reported having 3–8 years of experience; Six participants reported having 1–3 years of experience; Eight participants reported having no design experience. We categorized designers with over 3 years of design experience as expert designers and designers with less than 3 years of experience as novice designers, adding up to 10 expert designers and 14 novice designers. Six of the participants reported having less than six hours of sleep the previous night—a notable factor that could potentially affect neurological responses and mood states. The overall study protocol was approved by the Institutional Review Board prior to the start of research activities. 

\begin{figure}[h!]
  \includegraphics[width=\textwidth]{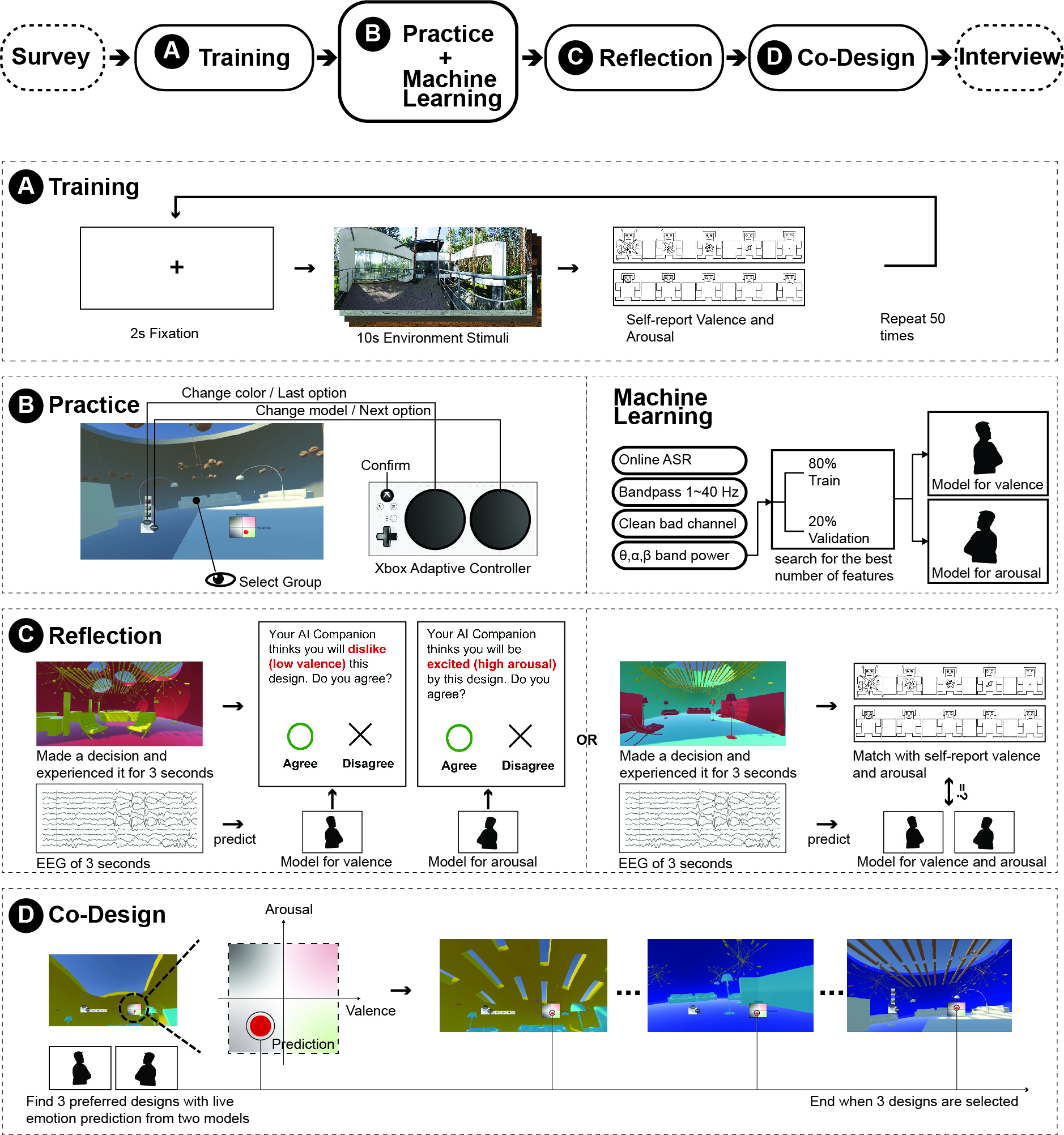}
  \caption{Schematic Overview of the Experiment Procedure: (A) Participants Experience Fifty Panoramic Environmental Stimuli and Self-Report Valence and Arousal; (B) Participants Practice Using Xbox Controller to Change Design Options; (C) Participants Express Agreement with "Multi-Self" and Report Valence and Arousal of the Selected Designs; (D) Participants Select Three Designs without Interruption with Real-time Emotion Predictions.}
  \Description{Protocol}
  \label{fig:Experiment}
\end{figure}

\subsection{Experiment Procedure}
The experiment consists of four main sessions (Figure \ref{fig:Experiment}). Sessions were conducted for one participant at a time. Each participant was first asked to complete a pre-experiment survey. The researchers then assisted the participant with fitting the EEG and VR equipment. The participant was asked to engage in a training session so that the BCI could become familiar with their individual emotional responses. This involved observing a panoramic VR image of an architectural environment for 10 seconds while we recorded EEG data and then completing the brief SAM scale \cite{lang1995emotion} to report arousal and valence. The process was repeated until 49 pre-selected environments were experienced, which took around 15 minutes in total.

Next, participants were placed into a practice session with limited design options to familiarize themselves with the head-pointing and Xbox adaptive controller controls. A research assistant gave instructions such as “please change the lobby envelope to a different model,” and the practice continued until the participant was able to fluidly use the controls. During this time, the researchers pre-processed the data collected during the training session and used it to complete the machine-learning classification. This process took around 5 to 10 minutes, and we explained to the participants that we were analyzing the data while they learned the design controls. 

Each participant then continued to the testing/validation session. In this part of the experiment, all of the lobby design options were available to the participant, and they were asked to experiment with different options to create a preferred design. We used two techniques to validate the accuracy of the feedback tool. First, when the participant made any change to the design, they were prompted to wait for 3 seconds, during which their brain activities were analyzed and classified by the BCI tool. The participants were then shown the neurological feedback and asked through a pop-up panel if they believed it to be accurate (Figure \ref{fig:Classify}). For example, if the model predicted the user was feeling high arousal, the popup window would say “Your AI companion thinks you would feel excited about this design change, do you agree? (Yes/No).”  After making another design change, the participant would experience the second validation method, which asked them to report arousal and valence via the SAM scale without seeing the neurological feedback. These two validation methods alternated until they had each occurred 6 times.

Finally, the participants were allowed to continue freely working on lobby designs until they were satisfied with the outcome, during which time they could view the ongoing feedback from the BCI response-tracking tool. Each participant was asked to complete three different lobby designs in this fashion, which generally took about 15 minutes. When satisfied with their work, the participants removed the VR and EEG equipment and completed a short debriefing interview about their experiences (Figure \ref{fig:Experiment}).

\subsection{Outcome Measures}
To gain a comprehensive understanding of machine learning performance, we evaluated it based on three distinct metrics: (1) offline validation accuracy, (2) online user agreement with the displayed neurological feedback, and (3) consistency between online self-reporting and unseen neurological feedback. Metrics one and three measure objective accuracy while metric two measures perceived accuracy. The purpose of this pilot study was to conduct proof-of-concept testing and obtain initial user feedback. Because our training data was imbalanced, we measured the imbalance ratio calculated by the division between the number of the class with the most samples and the number of the class with the least samples.

Validation accuracy was calculated as the average of 5-fold validation, which meant that we partitioned the data into five chronologically separate parts to avoid temporal correlations \cite{li2020perils}. We repeatedly used 80\% of the data as the training set to classify the other 20\%. In this process, the accuracy of the machine-learning outcomes could be evaluated by comparing the predictions against the actual remaining 20\% of the training data. Training accuracy refers to the proportion of correct predictions that a model makes on the training data set.

The user agreement validation was a simple calculation of the number of agreements divided by the total number of trials. Though this measure was subjective and susceptible to bias, it described perceived accuracy and thus helped us to understand the perceived trust between the user and the tool.

The consistency between online self-reports of arousal and valence vs. the unseen machine-learning predictions was also measured as the number of agreements divided by the number of total trials; these outcomes were a better indication of the objective accuracy rate.

Qualitative data were gathered during semi-structured exit interviews to understand the potential impacts of emotional feedback on metacognitive feelings (i.e. feelings of confidence). Our discussion was initiated by the following seven introductory questions: [Q1] Could you use a few words to describe your experiences in the design scene? [Q2] Do you think you built some trust with the AI system? [Q3] Do you think the real-time visualization of valence and arousal changed your decision-making? [Q4] Do you think the visualization of valence and arousal encouraged you to explore more design options? [Q5] What was the greatest challenge during your design process with the AI system? [Q6] If you had magic, what changes you wish to make for this design tool to aid your design process better? [Q7] What other design scenarios do you think this design tool can be applied to?

We also evaluated the user experience quantitatively using the User Experience Questionnaire (UEQ) \cite{laugwitz2008construction}. Participants were instructed to evaluate their experience of designing with real-time emotional feedback. The UEQ assesses the quality of interactive products and consists of six sub-scales. “Attractiveness scale measures the overall impression of the product; Perspicuity measures if the product is easy to understand and learn; Efficiency measures if the interaction is efficient and fast; Dependability measures if the user feels in control of the interaction; Stimulation measures if the product is exciting and motivating; Novelty measures if the product is innovative and creative.” \cite{schrepp2017construction}. The scale ranges from -3 to 3. The UEQ helps us understand if the design tool offers a decent user experience to the users in addition to subjective and objective accuracy. 

\section{Results}
\subsection{Machine Learning Classification}
\begin{figure}[h!]
  \includegraphics[width=\textwidth]{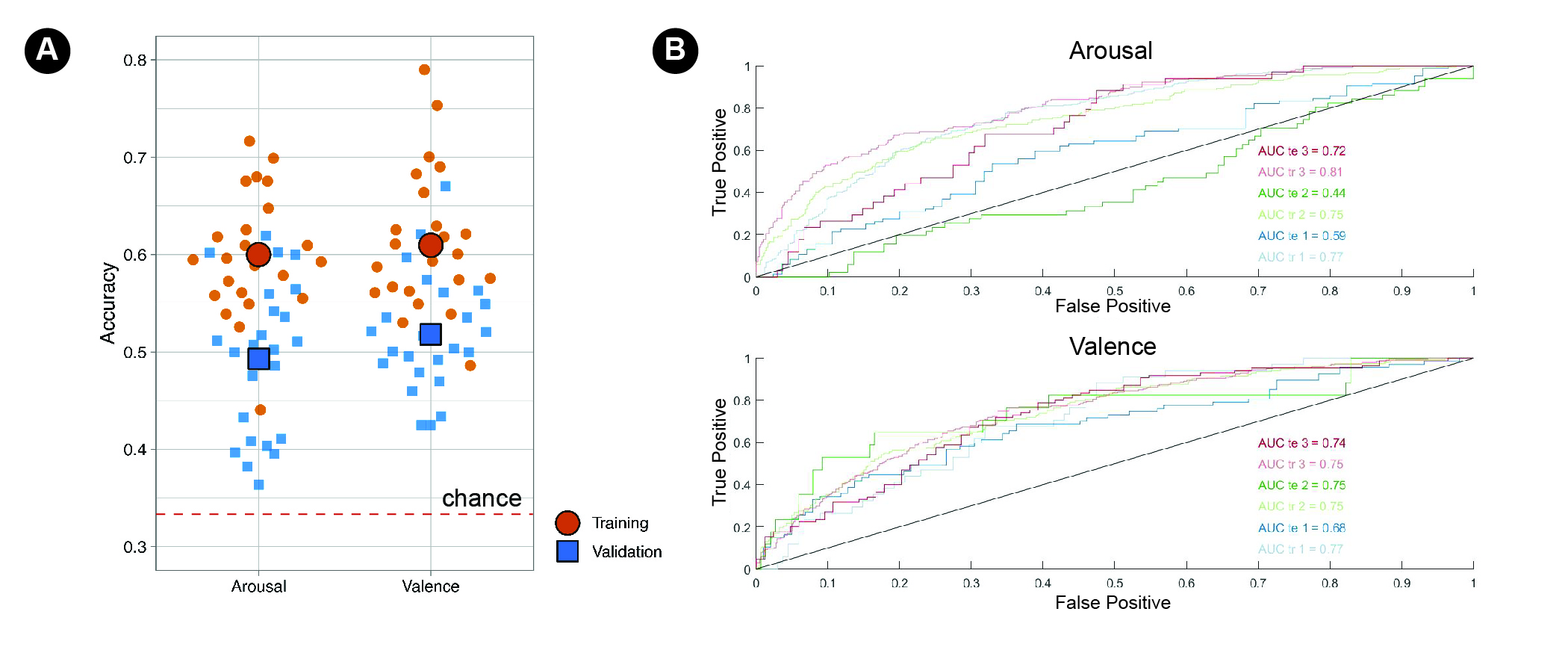}
  \caption{(A) Offline Training and Validation Accuracy of All Participants; and (B) ROC for Arousal and Valence Classification of Participant Two.}
  \Description{result_c}
  \label{fig:result_c}
\end{figure}

The average offline validation accuracy was 49.3\% (SD = 7.8\%) for the “arousal” dimension of affective response, and 51.8\% (SD = 6.0\%) for the “valence” dimension. This outcome was moderately strong since we were using a three-fold categorization schema (high, neutral, and low). The chance of randomly obtaining an accurate classification would be only 33\%. Achieving 49–52\% accuracy thus indicates a reasonably good predictive capability (Figure \ref{fig:result_c}). The average imbalance ratio for arousal is 2.71 (SD = 1.79). The average imbalance ratio for valence is 2.94 (SD = 1.44).

\subsection{Design Tool Validation}
\begin{figure}[h!]
  \includegraphics[width=\textwidth]{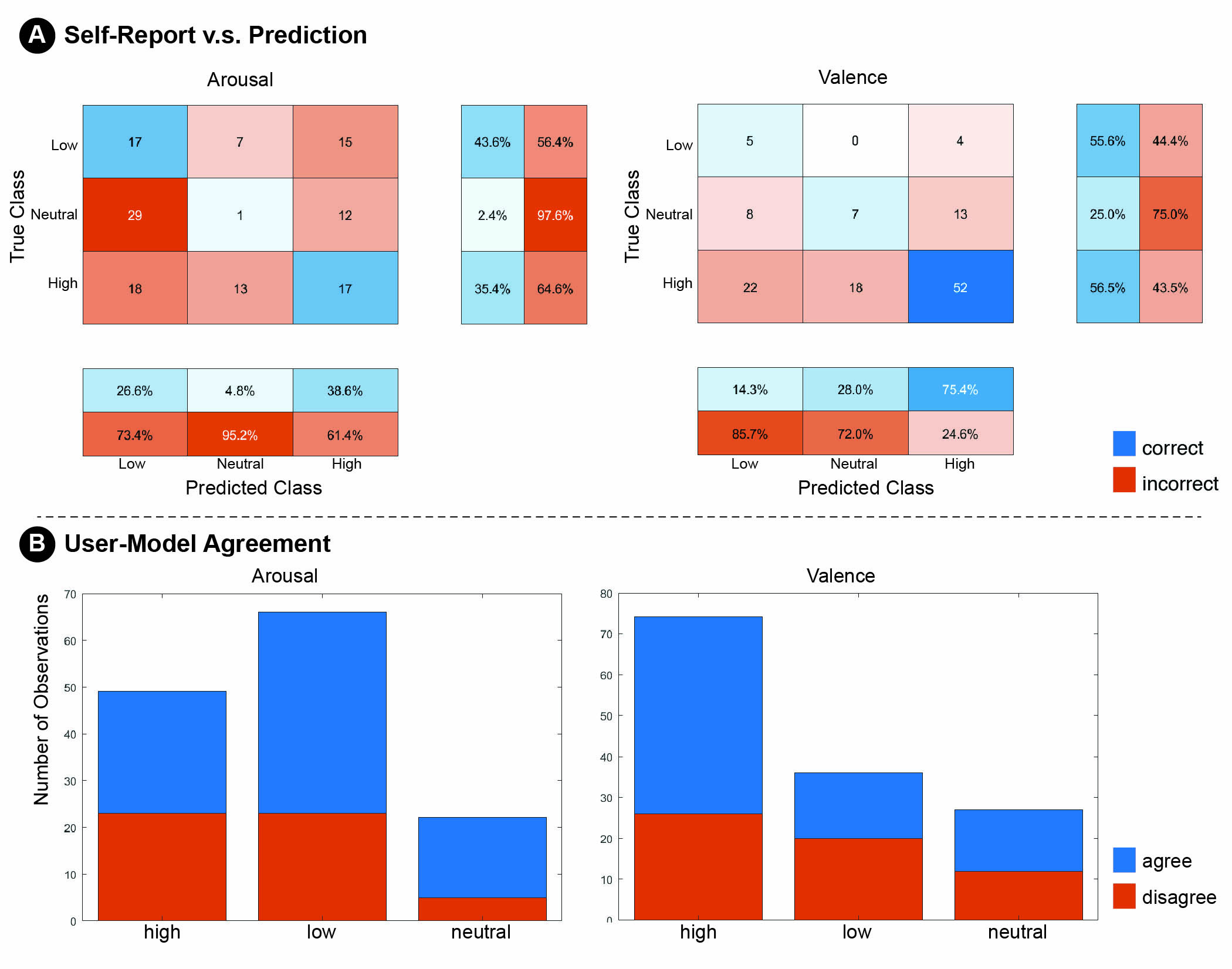}
  \caption{(A) Confusion Matrix Comparing Self-reported Emotion vs. Unseen AI Predictions of All Participants during Online Session C (B) Extent of Subjective User Agreement with the AI Feedback of All Participants.}
  \Description{result_i}
  \label{fig:result_i}
\end{figure}

In regard to the online validation, we collected a total of 137 responses for the user-agreement metric and 129 responses for self-reported measures vs. unseen predictions (Figure \ref{fig:result_i}). There were interesting discrepancies between these metrics for both valence and arousal. In the case of valence (positive or negative affect), participants agreed with the AI’s prediction 57.7\% of the time—however, their self-reporting matched unseen AI predictions 49.6\% of the time. In the case of arousal, participant agreement with the AI’s prediction was higher at 62.8\%, while their self-reporting matched unseen AI predictions only 27.1\% of the time. Our confusion matrix provides a detailed view of the performance of the classification model. As shown in Figure \ref{fig:result_i}, the precision and recall rates were calculated for each class. 

\subsection{Interview Feedback}
We used a qualitative content-analysis approach to parse the participants' responses from the exit interviews \cite{elo2008qualitative}. All responses were transcribed by the researcher, who also developed a preliminary coding schema based on commonly voiced themes related to the research variables. The two main themes identified in the data were: (1) attitudes and usage perspectives toward the BCI design tool and (2) expressions of uncertainty modulation and metacognitive awareness associated with the tool’s feedback. We identified responses related to metacognition if participants mentioned their reasoning and reflection processes. The same researcher and another research assistants independently reviewed the transcripts and coded the data segments by these themes. After completed their independent coding, they met to collaboratively resolve any disagreements about how the text segments were organized. The resulting themes and example quotations are presented in the following sections. 

\subsubsection{Attitudes and Usage Perspectives}
We identified five different attitudes towards the BCI design tool as expressed by participants, which in some cases led the participants to use the tool in a different manner. Some participants expressed more than one of these attitudes at different points in their interviews.

The first attitude was to treat the tool’s feedback as an important metric to guide design decisions. This outlook was expressed by 20.8\% (n = 5) of the participants. With this attitude, participants would continue testing different design options until the tool’s feedback moved in a direction they regarded as favorable. For example, Participant 16 said, \emph{“I would use it to see if the prediction changed. And if it changed, I would keep a design, even if I didn’t necessarily want to.”} Participant 18 said, \emph{“I chose things that gave me more positive [valence] and less arousal.”} Participant 11 said, \emph{“My emotion prediction tells me I am interested in it, and I thought about it and eventually did end up settling on some of them.”} These participants used the tool’s feedback as prompting to reconsider and evaluate their design decisions.

The second attitude was to use the system’s feedback as a confirmation of indigenous feelings. This outlook was expressed by 25\% (n = 6) of the participants. In these responses, participants indicated that they made a design decision first and then checked the model for potential confirmation. Participant 9 said, \emph{“Once I got to a design decision that I liked, I would then look at the quadrant. And if it aligned, then I would almost take that as a second voice telling me if that decision is right or not.”} Participant 13 said \emph{“I made the arrangement of the furniture, then I saw, like the dot move to the right direction, which I expect, then I think it’s good.”} Essentially, these participants were using the tool as a means to boost their confidence in their design skills.

The third attitude was to use the system as a final judgment when the participant was undecided. Only a small number of participants (8.3\%, n = 2) fit into this category. They indicated that they simply ignored the tool most of the time, and only checked for its advice when they had trouble selecting between two design options. Participant 15 said, \emph{“When I was really hesitating between two options, like when I’m not sure—oh, is this better? Or is this better? I could like look at it and be like, oh, it says I like this better. Okay, sounds good.”}

Many of the participants (33.3\%, n = 8) expressed an attitude of curiosity and playfulness when interacting with the tool. They regard it as an opinionated companion that could provide insights into design goals and prompt exploration. Participant 6 said, \emph{“I intentionally went against the predictions to see what happen.”} This participant added, \emph{“I would be happy to see if the AI has the same opinion as mine, but I still trust my current emotions.”} Participant 24 said, \emph{“I tried to cater to a few different emotions based on how it showed me I was feeling.”} 

Another large group of participants (33.3\%, n = 8) did not engage in any of the aforementioned attitudes, but instead ignored the tool’s feedback completely. One of the stated reasons for this reaction was that they felt confident in their emotions and did not feel they could gain any insights from the feedback. For example, participant 20 said, \emph{“I didn’t really look at [the affect feedback] ... I think I am fully aware of my feelings.”} Similarly, participant 22 said, \emph{“I didn’t pay much attention to it; I mainly design based on my own emotions. I focus more on how I feel by myself.”} Some participants ignored the tool’s feedback because they believed emotions were fleeting and should not be prioritized during architectural design. For example, participant 1 said \emph{“I felt designing a space would depend on many other factors, such as function.”} Some participants indicated that they perceived the feedback as an unhelpful and untrustworthy distraction. For example, participant 3 said, \emph{“I actually didn’t look at the emotion feedback that often because it didn't change that much.”}

\subsubsection{Metacognition}
In the interview data, participant expressions of metacognition were associated with three types of experience: (a) using the tool to continuously track one’s own emotional responses; (b) observing a mismatch between the tools feedback and conscious/indigenous feelings, and (c) observing a match between the tool’s feedback and conscious/indigenous feelings. Some participants expressed more than one of these attitudes at different points in their interviews. In total, 45.8\% (n = 11) of the participants mentioned some type of metacognitive reaction to the tool.

Observing continuous experience via the tool’s feedback encouraged the participants to view their designs from a third-person perspective. This experience was mentioned by 16.7\% (n = 4) of the participants. For example, participant 2 said, \emph{“It’s interesting enough to see the changing of my own experience.”} Participant 5 said \emph{“The prediction could turn the designer into the audience position, and think about what they would feel as the first-time audience.”} Participant 14 said \emph{“I actually didn’t feel my emotions that well, so the tool actually encourages me to reflect. It forces me to make a choice like it’s right or not.”}

The largest portion of metacognitive reflections was associated with a perceived mismatch between the tool’s feedback and indigenous feelings. This experience was mentioned by 41.7\% (n = 10) of the participants. Interestingly, the perception of mismatch led to a wide range of responses. Some participants were prompted by the feedback to reassess how they were feeling about the design. For example, participant 15 said, \emph{“I thought that something looks good, and it [the tool] was like, oh you don’t think it looks good. Intriguing.”} Some participants in this category were prompted by the tool to assess their conceptual design analysis. For example, participant 9 said, \emph{“I usually wouldn’t choose pure colors but now I wonder if there might be something in it that makes me like the design.”} another participant 16 said \emph{“I would look at design choices that I probably would not have made otherwise. I would look at the different, like brighter colors, even though I probably wouldn't have done that before.”} In general, we found that experiences of mismatch encouraged participants to explore a wider design space and consider alternative solutions.

In contrast, however, some participants felt that the tool had been trained “incorrectly” and wanted to revise it to better match their indigenous perspectives. Participant 21 said, \emph{“After the online session, I wish I could change how I answered in the training session so the model could serve my purpose better.”} A few participants rejected the tool outright. Participant 15 said, \emph{“I am like, man this color looks so bad, and [the tool] told me I would like this color. I didn’t trust it.”} Other participants indicated that they trusted the tool only when it provided certain types of feedback. Participant 11 said, \emph{“I found when the prediction moved pretty fast to the bottom right, I disagreed with it.”} The same participant added, \emph{“I trust the movement of the predictions rather than their positions in the quadrant.”}

When the participants mentioned that the tool’s feedback was a close match for their indigenous feelings, they associated this experience with an increase in confidence and the ability to narrow down the design space. This type of metacognitive experience was mentioned by 8.3\% (n = 2) of the participants. For example, participant 4 said \emph{"I think I'll make decisions a lot more, more efficient."} Another notable finding in this area was that many of the participants who discussed matching feedback from the tool also indicated that they only rarely referred to its output. For example, Participant 13 said, \emph{“It makes me confident about my decision but I do design based on my preference, so I didn’t pay much attention to it.”} A persistent pattern in this category of response was that the participants were using the tool when needed to improve their confidence, and otherwise did not give much consideration to its feedback.

\subsubsection{Acceptance Factors}

While perceptions of the tool’s accuracy were not associated with the extent to which participants used it, such outlooks did have an association with levels of acceptance toward the tool. Participants who reported higher levels of perceived feedback accuracy also provided more positive commentary during the interviews; for example, Participant 2 stated that, \emph{“Seeing the emotional responses motivated me to move the red dot to the top-right corner.”} In this category of response the participants frequently mentioned using the BCI tool in the intended fashion. Participant 24 mentioned that the tool prompted a wider design exploration: \emph{“I went through different design options to explore how the prediction went.”} Participant 9 emphasized improved confidence in their design choices: \emph{“the visualization helped me confirm my judgement”}.

In contrast, participants who felt the tool was inaccurate reported lower levels of trust and in some cases outright hostility. Participant 6, for example, said that the tool was trying to direct them away from designs that they actually liked and that they had to \emph{“choose against the AI’s decision.”} It was hard to determine causality in these types of responses—whether participants received mismatched feedback and thus came to distrust the tool, or whether initial distrust led them to focus on instances of mismatched feedback and vehemently reject it.

Another factor that emerged in the interviews in relation to the tool’s acceptability was the intuitiveness and detail level of the feedback visualization. Participant 7 brought up the question of BCI transparency, stating that, \emph{“It would be more convincing if we could see the raw brain activity data in some artistic way at the same time. If I could understand how those high-level predictions were generated, then I would trust the tool’s predictions more.”} Another suggestion from Participant 7 was that the biofeedback could be shown on an avatar model: \emph{“We were experiencing the architecture in a first-person perspective. Maybe you could sometimes change it to a third-person perspective so we can see the avatar of myself, and monitor that avatar’s biofeedback.”}

Finally, we observed that the extent of participants’ design experience was negatively associated with acceptance of the tool. expert designers are more suspicious of the design tool and tended to use the tool differently from novice designers. Among the 10 expert designers, 60\% (n = 6) expressed low trust in the tool and only two of them expressed high trust. In a contrasting manner, only 7.1\% (n = 1) expressed low trust and eight of them expressed high trust in the tool among the novice designers. Regarding how they use the tool, expert designers never treat the predictions as important metrics and four of them use predictions to confirm their ideas or motivate exploration. Four novice designers, however, treat the predictions as important metrics, and six use predictions to confirm or explore. Despite expert designers demonstrating much lower trust in the design tool, their metacognitive feelings emerge similarly to novice designers. 40\% (n = 4) expert designers and 57.1\% (n = 8) novice designers reported metacognitive feelings. 

\subsection{User Experience Questionnaire}
Regarding the user experience of designing with real-time emotional feedback, five sub-scales of UEQ were within the "above average" or "good" range based on the UEQ benchmark \cite{schrepp2017construction}: \emph{Attractiveness} (M = 1.22, SD = 0.81), Efficiency (M = 1.14, SD = 0.81), Perspicuity (M = 1.76, SD = 0.82), Stimulation (M = 1.32, SD = 1.01), and Novelty (M = 0.94, SD = 1.01). Dependability was rated as "below average" (M = 0.79, SD = 0.63). This lower score is not entirely unexpected given the tool's intended function to introduce an element of uncertainty into the design process. This uncertainty was intended to stimulate and challenge the designer's intuition and creativity, rather than to serve as a perfect predictive tool. In this context, the "below average" dependability rating could be the outcome of introducing uncertainty. Future studies could explore strategies to maintain this productive uncertainty while improving perceived dependability.

\section{Conclusion and Discussion}

In this research, we developed a closed-loop BCI-VR design tool to augment users’ real-time awareness of their affect levels based on their brain dynamics as recorded by EEG. The goal was to create a BCI application that could support design creativity by modulating feelings of confidence and stimulating metacognitive reactions. The results of the study can be discussed in three categories: the EEG classification results, user experiences in interactions with the system, the possible applications in the design process, and expanding the framework of existing creativity support tools and co-creative process.

In regard to the success of the EEG classification, the validation accuracy of our BCI tool could be improved but was in line with previous studies that focused on predicting valence and arousal based on EEG data for a 3-class classification \cite{dabas2018emotion, suhaimi2020eeg, marin2018affective, liu2017real}. This accuracy should also be considered in the context of the overall use situation—whereas prior studies in this area have focused on environments intended to elicit strong affective responses (for example, by displaying emotionally stimulating videos), we evaluated responses during an ordinary design task. Thus, the degree of accuracy achieved is notable within the current state-of-the-art of the field. However, while the EEG classification performed much better than the chance level, it should probably still be regarded as insufficient to support widespread satisfactory user experiences in real-life scenarios. In future work, there are a variety of options for potentially improving the tool’s performance, such as using more salient training stimuli, increasing the number of stimuli trials, and allowing users to conduct longer practice sessions. More advanced BCIs may eventually be able to adapt to individual users over a long duration of time, leading to increasingly accurate and helpful feedback as the user learns the tool and vice-versa. 

\begin{figure}[h!]
  \includegraphics[width=\textwidth]{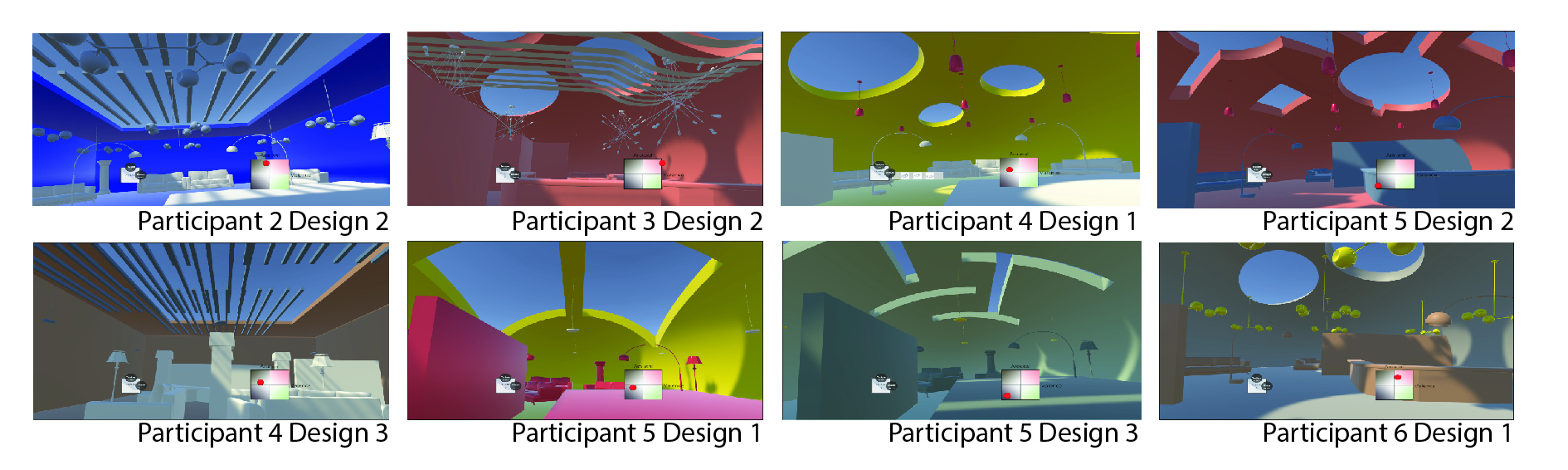}
  \caption{Design Outcomes Selected by Participants from Co-Design Sessions with "Multi-Self".}
  \Description{process}
  \label{fig:designOutcome}
\end{figure}

Compared to previous studies in this area that aimed to classify participants’ affect from EEG data, our work advances the application of BCI by developing a tool and visual feedback process that can be used to support creative processes in the context of design. Despite the accuracy limitations, it appears that for a considerable proportion of participants, the BCI tool might have potentially fulfilled its proposed role in fostering metacognitive evaluations and modulating the scope of uncertainty. We identified a variety of user behaviors and attitudes exhibited by different participants. While some users continuously examined the tool’s feedback to track their emotional responses during design, some participants relied on the tool only when they noticed a strong match or mismatch between the tool's feedback and their conscious evaluations. Some users intentionally tried to alter the BCI feedback through their design decisions, thereby using the tool as a mode of design exploration, while some participants relied on the tool primarily when they were faced with uncertainty or needed a confidence boost. Even for those participants who decided to reject the BCI’s feedback, the tool appeared to prompt reflection, promote risk-taking, and enhance confidence in the users’ ultimate design conclusions. Despite our initial assumption of participants favoring designs with high arousal and positive valence, the diversity of their attitudes resulted in design selections across a wide emotional feedback range (Figure \ref{fig:designOutcome}).

We found a wide range of individual differences in the tool’s performance and in attitudes toward the tool. More experienced designers are less likely to trust or appreciate the tool’s feedback. It seems likely that this is a result of experienced designers having greater innate confidence in their design evaluations and more effective intuitions, making them rightly question the accuracy of the tool, but also possibly due to being more grounded in their careers and more accustomed to the positive reception of their work. Nonetheless, we found more metacognitive monitoring from experienced designers when the BCI design tool disagreed with them. The more positive responses received from novice designers implied a specific potential utility in educational settings. As BCI technology continues to expand and improve, it is likely that we will discover additional customization options and modalities for identifying individual neural responses to design and for visualizing those responses back to users.

The findings of this study advance the use of BCIs in a novel direction, taking a new step in the application of this technology to creative design context. While this research is still preliminary, we believe that BCIs have tremendous potential in design fields, both as a means to enhance the creative design process and in applications for testing the impact of design variables (e.g., evaluating user responses to a mentally relaxing/restorative design to improve its effectiveness). If expanded to multiple simultaneous users, the Multi-Self approach could help designers to identify collective responses and response ranges, and thus more effectively evaluate a design across a larger user base. The resulting data about the impacts of architectural design features could enhance the state of knowledge in fields such as environmental psychology as well as improve design outcomes in numerous contexts (urban exteriors, residential design, office design, video game design, etc.).

In contrast to existing creativity support tools that primarily enhance human perception and cognition \cite{rosenholtz2011predictions}, our work explores a fresh possibility: to enhance metacognition in the creative process, an area not deeply addressed in current research. Previous studies have predominantly focused on the various forms and modalities of biofeedback, with a central aim of improving mindfulness \cite{haghighi2020self, de2018augmented, yu2016livingsurface, roo2016inner}. Going beyond these approaches, we investigated a new potential application of neurofeedback systems: the integration of emotions into the design process. Introducing the concept of "co-design with myself", we suggest a potential method to augment the interactive creative design process by employing machine learning. This scenario, in our view, allows for a more personalized, self-reflective design experience, and aims to elevate the creator's inherent abilities. We humbly offer these ideas, hoping to spark conversations and encourage more research on enhancing metacognition in the field of creativity support tools.

\subsection{Implication for Design Practice}
This study contributes to a growing collection of incipient work that is focused on applying BCI as a modality of human–computer interaction during design scenarios \cite{roo2016inner, huang2017brain, kovacevic2015my, shankar2014human}. We expect that the impact of this research direction will continue to expand in the near future, as the BCI field in general is rapidly developing enhanced computational models and more sensitive hardware with improved signal quality \cite{stegman2020brain, gao2021interface}. As researchers look into the various potential applications of BCI for design, we recommend that affective computing has some of the strongest potential. Studies have shown that emotions play a crucial role in the design process and are central to the ways in which users experience built environments \cite{bhandari2019understanding, coburn2020psychological}. The application of BCI to capture and visualize emotional responses in real-time can make a strong contribution to better understanding this aspect of design, aiding the workflow of designers, and producing more innovative and effective design outcomes. Many designers, especially novices, may experience emotional stress and anxiety during the design process \cite{bhandari2019understanding}, and the MultiSelf tool presented here has the potential to help designers mitigate and manage these reactions through modulating subjective uncertainty. In the broader picture, such tools have tremendous potential in evaluating user responses to design and possibly even developing architectural designs that automatically adjust to users’ affective states. 

\subsection{Limitations and Future Work}

Our initial user studies suggest that the real-time depiction of users' emotions could potentially boost metacognitive monitoring. However, given the current unsatisfactory classification accuracy of the tool and the lack of participants' awareness of this accuracy, these findings should be cautiously interpreted. Importantly, a tendency for novice designers to place higher trust in the tool was observed, which underscores potential risks for misguidance if the feedback tool consistently misjudges their emotions. Future studies ought to consider notifying users about the prediction accuracy in usability studies.

Future studies should incorporate behavioral data into their analyses. For instance, researchers could investigate whether changes in the visual feedback of predicted emotions lead to increasing changes in design options, or if specific emotional feedback leads to pauses, which could be an indicator of the reflection process. The objective behavioral data could provide stronger evidence of the design tool's performance besides classification accuracy.

Despite the large range of design options that we presented to participants in this study, the framework of the lobby design task was less complex than truly open-ended real-world scenarios. It is possible that this simplified task may have been unable to incite a high degree of subjective uncertainty, making participants less motivated to use the BCI tool for guidance. In future work, we will seek to evaluate the use of the MultiSelf tool in more challenging design contexts, which can be done by expanding the design possibility space and setting up more complex design constraints.

Another possibility for future work in this area is to expand the included biofeedback beyond EEG data, by integrating metrics such as skin conductance, heart rate, and facial expression recognition into the classification process. This may assist in increasing the robustness of the BCI’s feedback, and make it more effective for a wider range of participants. It may also be beneficial to visualize the biofeedback to the designers in a more intuitive fashion, for example by replacing the moving dot on a coordinate plane with ambient feedback such as haptic stimulation, lighting changes, or sound changes. Some participants suggested using a more personified visualization such as an avatar or photo of the user to present the visualization. Rather than having to consciously track and interpret such feedback on a coordinate display, a smoother assistive system could be designed to gently notify the designer when there is a positive or negative response. For some users, it may be beneficial to include more advanced analytical options such as graphs of changes in affect over time and more information about the raw classification data.

Finally, this pilot study was limited in the quantity and diversity of participants. Many of our participants were experienced designers, who might have relatively low levels of uncertainty or reduced need for reflective assistance. Focusing on novice designers in future research would be in line with the positioning of MultiSelf as an educational or early-career feedback tool. This could be done in conjunction with broadening the diversity of the participant sample and collecting more data about demographic variables as related to user responses. In the broader picture, we are also considering the use of reflective design assistance as a means of democratizing design, providing enhanced user customizability for design products, and encouraging human/AI collaborations that can expand the horizons of the field.

%%
%% The acknowledgments section is defined using the "acks" environment
%% (and NOT an unnumbered section). This ensures the proper
%% identification of the section in the article metadata, and the
%% consistent spelling of the heading.
\begin{acks}
Many thanks to Jesus G. Cruz-Garza for his supportive advice of developing the BCI.
\end{acks}

%%
%% The next two lines define the bibliography style to be used, and
%% the bibliography file.
\bibliographystyle{ACM-Reference-Format}
\bibliography{main}
%%
%% If your work has an appendix, this is the place to put it.
%\appendix

%\section{Appendix}

%\subsection{Part One}

%Can discuss detailed EEG data processing protocol here.

\end{document}